\documentclass[10pt,fleqn,twocolumn]{revtex4}
\usepackage[dvips]{graphicx}
\usepackage[usenames,dvipsnames]{color}
\usepackage{amsmath}
\usepackage{amssymb}
\usepackage{amsfonts}

\newcommand{\ket}[1]{|#1\rangle}
\newcommand{\bra}[1]{\langle#1|}

\begin{document}

\title{Observation of room-temperature spontaneous superradiance\\
 from single diamond nanocrystals}
\author{Carlo Bradac$^{1,2,\ast}$, Mattias Johnsson$^{1,2,\ast}$, Matthew van Breugel$^{1,2}$, Ben Baragiola$^{1,2}$, Rochelle Martin$^{1,2}$, Mathieu L. Juan$^{1,2}$, Gavin Brennen$^{1,2}$, and Thomas Volz$^{1,2}$}
\affiliation{$^1$ Department of Physics \& Astronomy, Macquarie
University, NSW 2109, Australia} \affiliation{$^2$ ARC Centre of
Excellence for Engineered Quantum Systems, \\Macquarie University,
NSW 2109, Australia}
\thanks{These authors contributed equally to this work.}
\maketitle

{\bf Superradiance (SR) is a cooperative phenomenon which occurs when $N$ quantum emitters couple collectively to a mode of the electromagnetic field as a single, massive dipole moment that radiates photons at an enhanced rate. The conditions required for SR arise from the \emph{indistinguishability} of the emitters with respect to the field mode. As set forth by Dicke in his seminal 1954 paper \cite{dicke_coherence_1954}, spatial indistinguishability occurs when the emitters are confined to a volume much smaller than the scale set by the wavelength of the emitters' optical transition, $V \ll \lambda^3$. Additionally, the emitters must be spectrally indistinguishable, which complicates the study of SR in a solid-state setting due to large inhomogeneous broadenings and unavoidable dephasing. Previous studies on solid-state systems either reported SR only close to liquid-helium temperatures, and/or from sizeable crystals with at least one spatial dimension much larger than the wavelength of the light \cite{cong_dicke_2016}. Here, we report observations of room-temperature SR from single, highly luminescent diamond nanocrystals with spatial dimensions much smaller than the wavelength of light, and each containing a large number ($\sim 10^3$) of embedded nitrogen-vacancy (NV) centres. After excitation of the nanodiamonds (NDs) with an off-resonant, green laser pulse, we observe i) ultrafast radiative lifetimes (LTs) down to $\sim 1$~ns, and ii) super-Poissonian photon bunching in the autocorrelation function of the light emitted from the fastest NDs. We explain our findings with a detailed theoretical model based on collective Dicke states and well-known properties of NV centres. Using a minimal set of fit parameters, the model captures both the wide range of different LTs and the nontrivial photon correlations found in the experiments. The results pave the way towards a systematic study of SR in a well controlled, solid-state quantum system at room temperature. Ultimately, quantum engineering of SR in diamond has the potential for advancing applications in quantum sensing, energy harvesting, and efficient photon detection \cite{higgins_superabsorption_2014}.}

The occurrence of SR, or cooperative emission, in an ensemble of identical emitters indicates the build-up of large-scale coherence between individual dipoles. Due to the many available pathways for photon emission from a system of $N$ indistinguishable initially excited emitters, the de-excitation process itself leads to the formation of highly entangled symmetric superposition states, so-called {\it Dicke states}. Dicke described the system of $N$ dipoles (or two-level emitters) using collective pseudospin operators with total spin $J=N/2$ and projection $M$, corresponding to $J+M$ excitations. He calculated the fluorescence rate $\gamma_{J,M} = \gamma(J(J+1)-M(M-1))$ with $\gamma$ being the single-dipole emission rate. As the system cascades down the ’Dicke ladder’ of states, the photon emission rate scales at maximum as $N^2$, an enhancement by a factor $N$ over $N$ independent dipoles, hence the name \emph{superradiance}. Observed initially in well-isolated, controlled laboratory settings \cite{skribanowitz_observation_1973}, SR has over time found applications in a variety of fields. For instance, it has been evoked as an underlying mechanism for exciton delocalisation in light-harvesting complexes \cite{monshouwer_superradiance_1997}. In astrophysics, SR is predicted to occur in the vicinity of black holes \cite{putten_superradiance_1999}, and in the field of precision metrology, a novel superradiant laser source was realised promising unprecedented narrow linewidths \cite{bohnet_steady-state_2012}. Further, the presence of highly entangled multi-particle states is an attractive prospect for quantum metrology \cite{zhang_quantum_2014}. Generally, Dicke states (or more precisely the $J$-subspaces) are immune to certain types of environmental noise that affect all emitters in the same way and cannot resolve individual dipoles \cite{gross_superradiance:_1982}. In a solid-state setting, this includes global dephasing due to long-wavelength phonon modes. However, local dephasing mechanisms, such as coupling to short-wavelength phonons and coupling to electric fields arising from ionization or other local defects, can have a detrimental effect on the cooperative behaviour of a system. Indeed, it is the simultaneous requirement of high spin density and low local decoherence that has made SR challenging to observe at room temperature in solid-state or atomic systems.

\begin{figure}[ht!]
    \centering
    \includegraphics[width=\columnwidth]{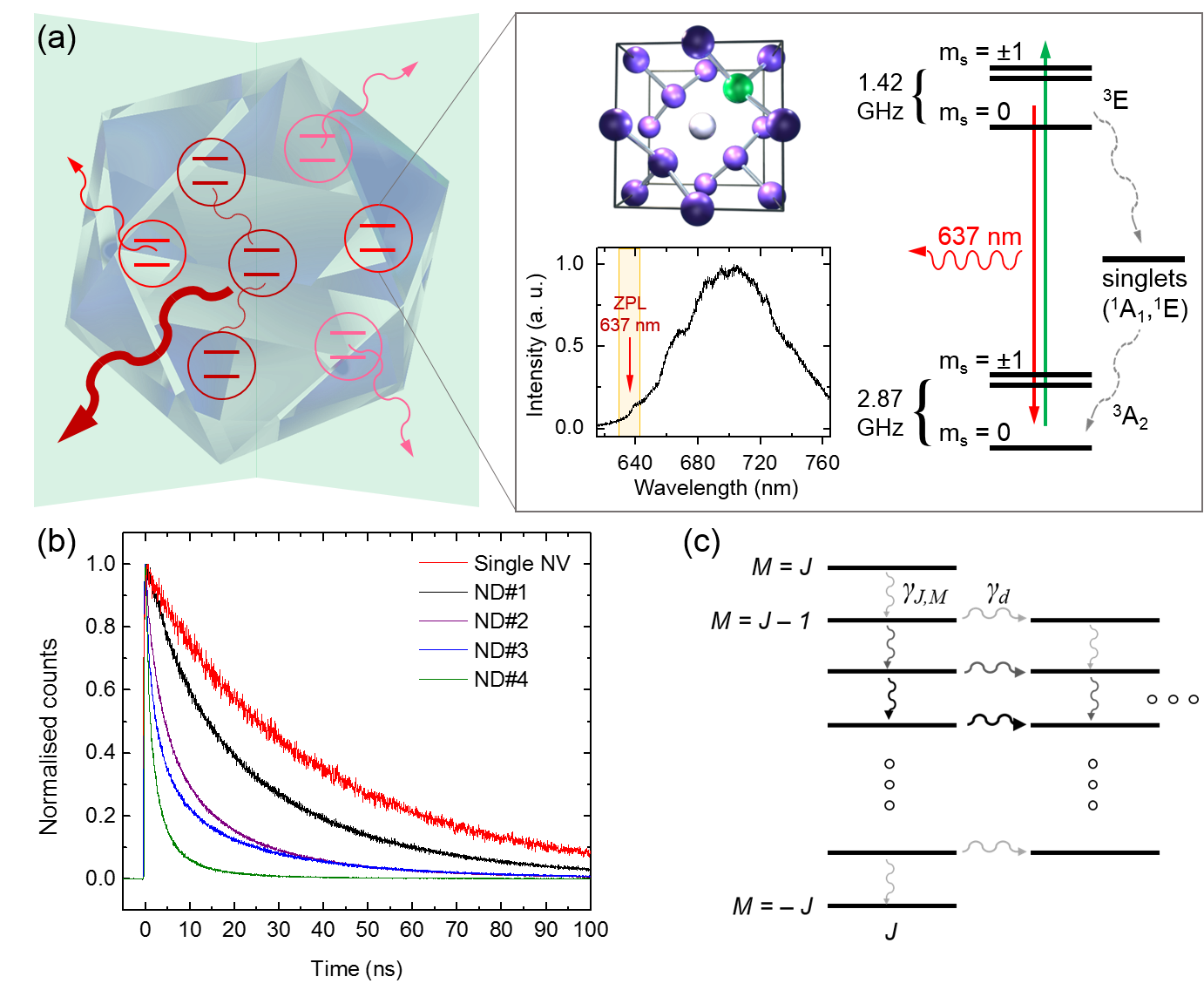}
    \caption{\textbf{Superradiance from nanodiamonds with many NV centres $\mid$} \textbf{a)} Graphic representation of cooperatively interacting NV centres emitting a superradiant burst from a nanoscale diamond crystal. The zoom-in shows the underlying crystalline structure around a single NV centre, with the substitutional nitrogen atom indicated in green and the vacancy in white. It also displays the level structure of a single NV centre \cite{doherty_nitrogen-vacancy_2013} and the corresponding fluorescence spectrum. Due to strong vibronic sidebands, only a fraction of photons is emitted into the ZPL at 637~nm. \textbf{b)} Measured normalised fluorescence decay curves for five different NDs, with LTs ranging from the usual few tens of nanoseconds for a single NV centre in a ND (red trace), to LTs around 1~ns for high-density NV NDs (green trace, ND $\#$4). \textbf{c)} Illustration of the Dicke ladder of states: Collective optical decay couples descending states within each pseudospin $J$-subspace at a characteristic rate $\gamma_{J,M}$. Local dephasing, at rate $\gamma_d$, decouples individual spins from the collective subspace, leaving the remaining spins in a smaller $J$-subspace. Thicker/darker decay lines denote stronger decay rates with maximum decay near $M=0$ states.}
    \label{fig:Intro}
\end{figure}

The NV centre (Figure \ref{fig:Intro}a) is an extrinsic diamond defect where two adjacent carbon atoms in the lattice are replaced by a substitutional nitrogen atom and a vacancy \cite{doherty_nitrogen-vacancy_2013}. Its most stable form, the negatively charged NV$^-$, displays triplet electronic ground ($^3A_{2}$) and optically excited ($^3E$) states, and intermediate singlet states ($^1A_{1}$ and $^1E$). The separation in energy ($ 1.945\,\rm{eV}$) between the ground and the excited states ($^3A$ --– $^3E$) corresponds to a zero phonon line (ZPL) at $637\,\rm{nm}$ followed by characteristic phononic sidebands associated with local vibrational modes \cite{davies_optical_1976}. These local vibrational modes are due to deformations in the lattice within a few unit cells of the defect and are characterized by ultrafast femto/picosecond, non-radiative relaxation. Previous work \cite{huxter_vibrational_2013} suggested that these local modes decay into global, long-wavelength, acoustic phonon modes which exhibit decay on a much longer timescale of a few tens of picoseconds, which in turn is still much shorter than any optical rate in the system. The decay of local into global phonons erases any information that the local environment would have gained and therefore ultimately preserves the coherence amongst the emitters and enables the subsequent superradiant photon emission (cf. Supplementary Information).

From an experimental point of view, the most salient feature of SR is accelerated optical emission, whose intensity burst can scale faster than linearly with the number of emitters. To investigate this phenomenon, we measured fluorescence decay of 100 separate NDs hosting a high density of NV centres (${\sim} 3 \times 10^6$ NV centres per $\mu$m$^3$, see Materials and Methods) and compared the decay curves against our theoretical model (see below). In addition, we measured brightness and size for the 100 NDs, and performed saturation-intensity and size-reduction measurements.
Figure 1d shows a subset of NV decay curves representative of NDs of different size and brightness. The red curve is the decay curve for a single NV centre used for reference. For some of the recorded fluorescence curves we observed LTs around 1~ns or even below, never reported before for NV centres (e.g. Fig. \ref{fig:Intro}c, ND$\#$4). Note that the $1/e$-LTs were extracted by fitting a standard exponential decay to the first nanosecond of the decay curve.


The theoretical model used to fit the LT curves is described in detail in the Supplementary Information. Briefly, we assume a collection of individual spectral domains (most likely corresponding to spatial domains within the nanocrystal), each containing a different number of NV centres that initially act collectively. The $m_{s}=0$ and $m_{s}=\pm 1$ populations are treated as two separate collections of domains. The inter-system crossing mixes the spin state populations, but this process is non-radiative and serves only to decrease the overall collective radiation. The collective behaviour breaks down over time due to local dephasing, projecting the collective centres partially into a lower dimensional collective subspace and partially into the non-collective space that undergoes standard exponential decay.

The inputs to the model are: {\it i)} the number $N$ of NV centres in each domain, {\it ii)} the initial state of each domain, {\it iii)} the underlying bright (radiative) and dark (non-radiative) decay rates, as well as {\it iv)} the local dephasing rates for the $m_{s}=0$ and $m_{s}=\pm 1$ populations. The dark decay rates are not well known; we take the best estimates from \cite{doherty_nitrogen-vacancy_2013} and use $2\pi\times 1.8$\,MHz and $2\pi\times 9.4$\,MHz for the $m_{s}=0$ and $m_{s}=\pm 1$ rates, respectively. The bright decay rates are heavily influenced by the size and geometry of each individual ND, meaning we cannot simply use bulk rates. To obtain the bright rate for each ND we perform an exponential fit on the long-term tail of the LT curve, well after the collective processes have ended, including the dark decay rates. This leaves the local dephasing rates, the number of centres in each domain, and the initial state of the collective space as free parameters in the model. To further constrain the model, we assume that the distribution of the number of NV centres is Gaussian across the domains and that the initial state consists of having each $M$-level in the symmetric Dicke ladder equally populated (Figure \ref{fig:Intro}d). This initial state assumption is not critical, as different distributions across the $M$-levels can provide equally good fits by making small changes to the mean of the Gaussian number distribution.

\begin{figure}[h!]
    \centering
    \includegraphics[width=\columnwidth]{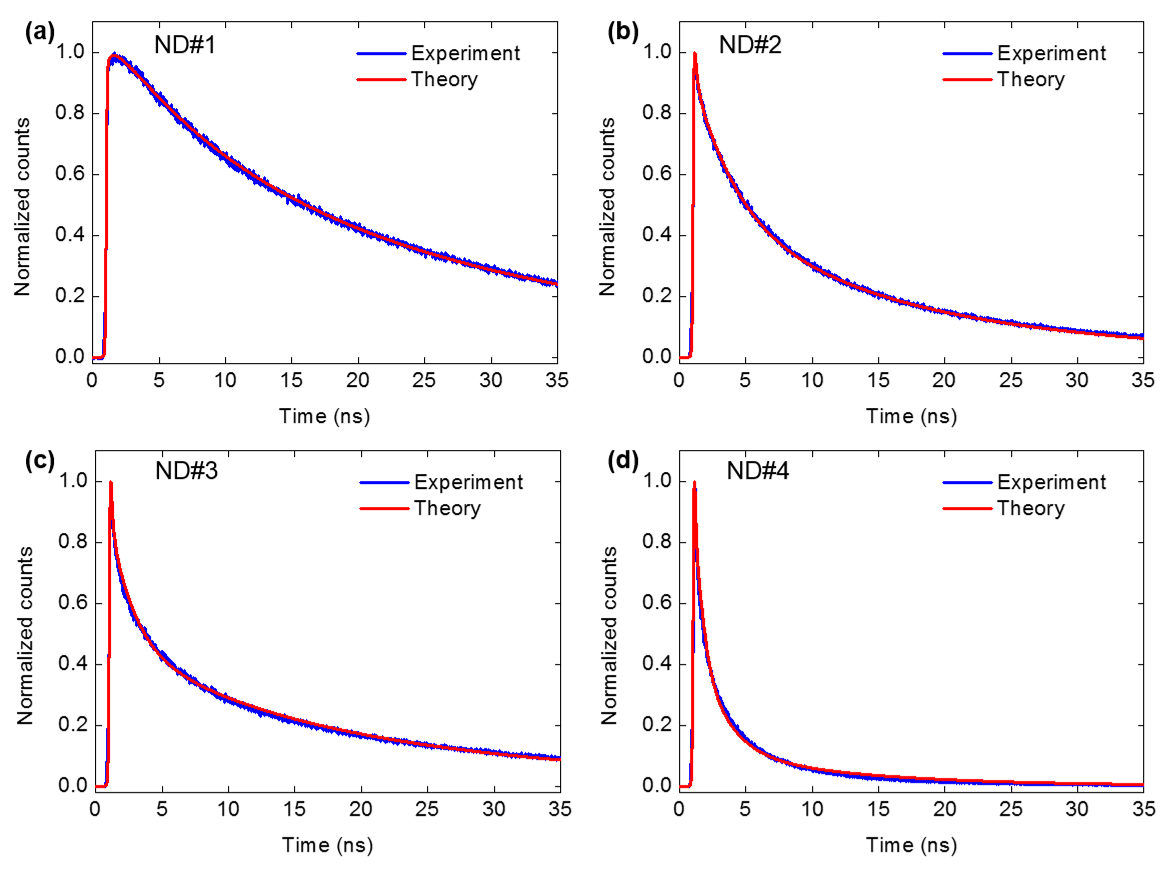}
    \caption{\textbf{Fluorescence decay curves and corresponding fits for four different NDs $\mid$} \text The four graphs display measured fluorescence decay curves
   [blue] with corresponding fits [red] obtained from our model (see main text), showing excellent agreement. Note that the curves are normalized to their respective maximum. The different NDs exhibit increasingly faster photo-emission, with corresponding LTs of $\{25, 3.6, 2.2, 1.1\}$~ns for ND$\#$1--4. The shorter LTs correspond to larger collective domain sizes of $N = \{2, 7, 10, 50\}$, respectively. For a quantitative comparison see the Supplementary Information.}
    \label{fig:LTs_fitted}
\end{figure}

With these assumptions, we find an excellent agreement between the fits from our model and the LT curves of each of the 100 NDs we characterized (see Supplementary Information). Figure \ref{fig:LTs_fitted} shows the fits for four NDs (ND$\#$1--4) representative of four distinct typical decay rates, each corresponding to a different collective-domain size with faster decay indicating larger domain size. The local dephasing rates extracted from the fits were largely consistent across all the NDs and varied between $\gamma_d^0/2\pi \sim 20$--$40$\,MHz and $\gamma_d^{\pm 1}/2\pi \sim 300$--$450$\,MHz for the $m_{s} = 0$ and $\pm1$ domains, respectively. We attribute the roughly ten times higher dephasing rates for the $m_{s}=\pm 1$ states to inhomogeneous electric fields throughout the crystal (see Supplementary Information). The fits also allow for the extraction of the initial NV spin polarization, i.e. the fraction of spins initially in the $m_s=0$ state. This ratio varied in the range ${\sim} 50$--$60\%$ across the investigated NDs and is in line with previous measurements of spin-polarization in high-density NV samples \cite{felton_hyperfine_2009, acosta_diamonds_2009}.

\begin{figure}[h!]
    \centering
    \includegraphics[width=\columnwidth]{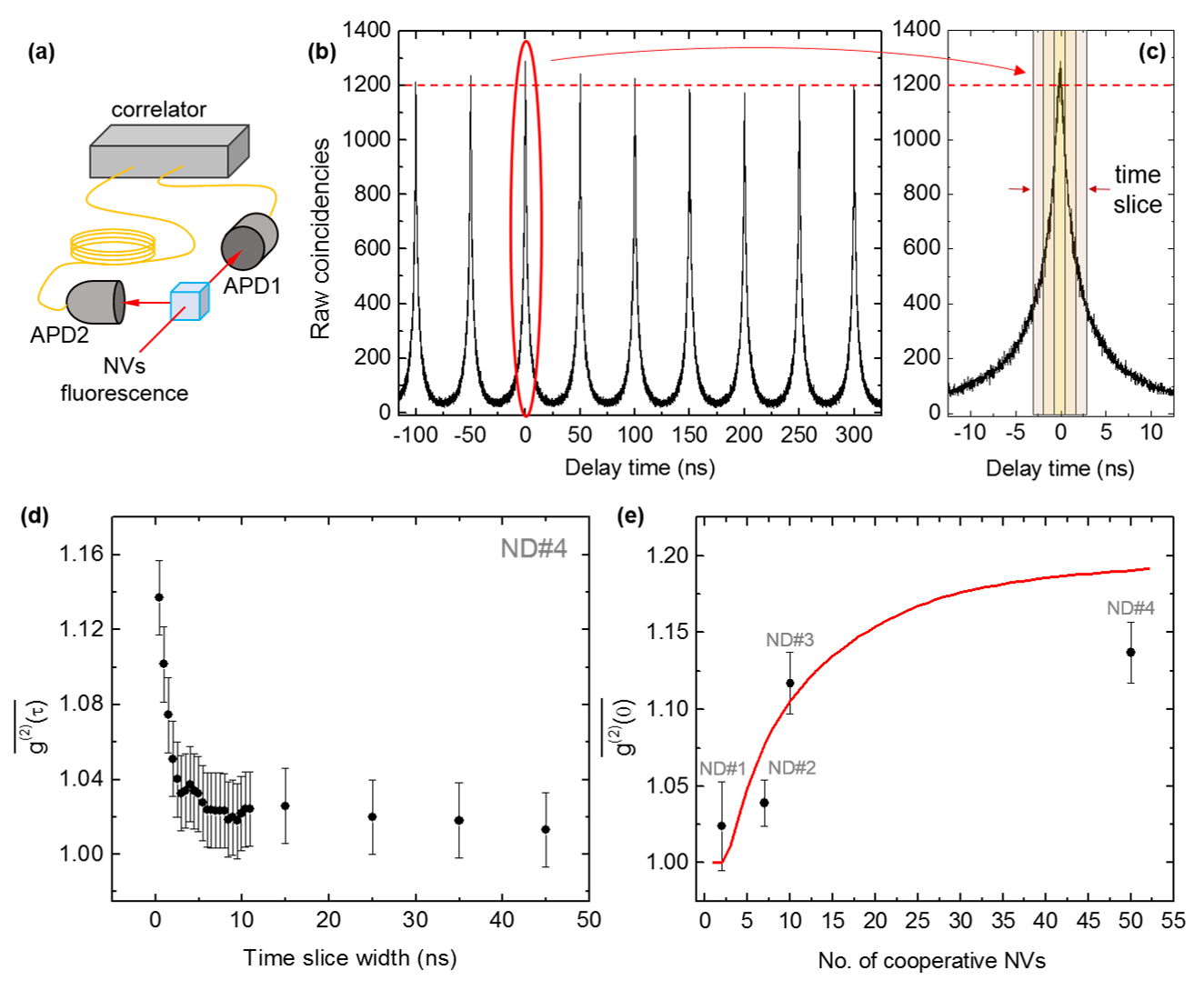}
    \caption{\textbf{Autocorrelation measurements $\mid$} \textbf{ a)} Schematic representation of the Hanbury-Brown and Twiss interferometer. \textbf{ b)} Normalized coincidences for ND$\#$4. \textbf{ c)} Time-slicing employed to evaluate $\overline{g^{(2)}(\tau)}$ in \textbf{(d)}. \textbf{ d)} Measured time-integrated autocorrelation function $\overline{g^{(2)}(\tau)}$, which approaches $g^{(2)}(0)$ as $\tau \rightarrow 0$. For ND$\#$4, $\overline{g^{(2)}(0)}$ crests at $1.14 \pm 0.02$ for a time-slice width of $0.5\,\rm{ns}$; it drops considerably as the width increases above $2-3\,\rm{ns}$ (after which the SR burst has exhausted) to then converge to Poissonian/random photoemission $\overline{g^{(2)}(0)} {\sim} 1$ expected from many NV centres at long times. Error bars are determined from the standard deviation of the area under the peaks, for each set of time slices, excluding the '0' peak. \textbf{ e)} Measured maximum value of $\overline{g^{(2)}(0)}$ for ND$\#$1--4, and corresponding theoretically estimated number $N$ of NV centres acting cooperatively to produce such value of $\overline{g^{(2)}(0)}$ using the initial state assumed by our model. The continuous line [red] sets the upper limit of our theoretical prediction.}
    \label{fig:Figure3}
\end{figure}

For the majority of the NDs, we found a typical cooperative domain size of $N \sim $1--2, indicating absent or very little collective behaviour. However, the faster decaying diamonds (e.g. ND$\#$2--4) were accurately fitted by using a higher mean number of centres acting collectively ($N\sim $10--50), as shown in Figure \ref{fig:Figure3}a--d. We attribute the lack of collective behaviour in the majority of our NDs to the material preparation method (see Materials and Methods), with the high-dose proton irradiation process followed by annealing yielding a high degree of spatial and spectral distinguishability amongst NV centres within the same ND host. However, stochastic variation gives rise to the existence of a few NDs exhibiting domains with large numbers of spatially and spectrally identical NV centres which do act cooperatively.

It should be noted that previous studies reported a decrease in the LT of NVs for centres produced via low-energy He-ion irradiation, with the decay time decreasing for increasing ion doses. This effect has been attributed to increased damage in the crystal lattice which provides nonradiative decay paths with faster dynamics \cite{mccloskey_helium_2014, orwa_raman_2000}. This is however inconsistent with our observations where we found that higher peak fluorescence correlated to faster decay rates (see Supplementary Information) -- the exact opposite of what would be expected if the shortening of the LTs was indeed due to non-radiative, dark pathways. In order to test quantitatively against other possible explanations for the observed fast decay dynamics, we also attempted to fit the observed lifetimes with both a bi-exponential and a deformed exponential \cite{gatto_monticone_systematic_2013} lifetime curve both of which gave clearly worse results.

To collect further experimental evidence for the validity of our theoretical model, we performed autocorrelation measurements (cf. Materials and Methods) by means of a Hanbury-Brown and Twiss interferometer (Fig. \ref{fig:Figure3}a) \cite{Jahnke:2016tg}. To ensure spectral indistinguishability of the photons we only analyzed the light from a narrow emission band around the ZPL (compare Figure \ref{fig:Intro}a). The measured time-integrated autocorrelation function $\overline{g^{(2)}(\tau)}$ revealed photon bunching for zero time-delay ($\tau \rightarrow 0$) for the fast-decaying ND NV centres, indicating super-Poissonian statistics. We measured values of $\overline{g^{(2)}(0)} > 1$, and as high as $1.14 \pm 0.02$ (Fig. \ref{fig:Figure3}b--e). Note that $\overline{g^{(2)}(0)}$ corresponds to the usually quoted $g^{(2)}(0)$. The value of $\overline{g^{(2)}(0)}$ is dependent on the initial state of the system, which in turn is determined by the preparation process.  The type of super-Poissonian photon statistics we obtained is consistent with the assumed initial state in our model (details see Supplementary Information). Figure~\ref{fig:Figure3}e shows the measured value of $\overline{g^{(2)}(0)}$ for the same four representative NDs (ND$\#$1--4) already analysed in Figure~\ref{fig:LTs_fitted}, plotted against the corresponding number $N$ of NV centres acting cooperatively as predicted by our theoretical model; the continuous line shows the upper limit of our theoretical prediction for perfect autocorrelation measurements. The good agreement between theory and experiment supports our theoretical model and complements the LT data. Together, they unambiguously demonstrate the underlying cooperative nature of fluorescence decay of the fastest NDs in our sample.

Our observation of SR in a true nanoscale, room-temperature solid-state system paves the way for a wealth of novel research directions. Immediate subsequent experimental steps include low-temperature studies for obtaining spectral information, and accessing spatial information through e.g super-resolution techniques such as stimulated emission depletion (STED) spectroscopy \cite{arroyo-camejo_stimulated_2013}. An obvious extension of our theoretical model incorporates the effect of dipole-dipole interactions which are expected to partially break the cooperativity amongst NV centres \cite{friedberg_limited_1972} but at the same time could allow for super-absorption~\cite{higgins_superabsorption_2014}. In addition, the NDs studied here are a novel system for exploring cooperative atomic forces in the context of optically trapped nanoparticles \cite{juan_observation_2015}. Alternative diamond colour centres, such as silicon-vacancy centres \cite{Vlasov:2014kl, rogers_electronic_2014}, exhibit a much smaller spread in transition frequencies and much-reduced phononic sidebands -- both indicate the potential for greater SR compared to NV centres. Incorporating colour centres in diamond into microscopic optical cavities \cite{hunger_fiber_2010} might allow for the observation of a solid-state analogue of the Dicke phase transition previously observed with cold atoms \cite{baumann_dicke_2010}. Finally we point out that deterministic implantation techniques \cite{mclellan_patterned_2016} and sophisticated material engineering in diamond could enable the controlled creation of mesoscopic ensembles of colour centres in a given spatial arrangement and with appropriately engineered photonic \cite{burek_free-standing_2012} and phononic environments \cite{rath_diamond-integrated_2013}. Colour centres in diamond, and more specifically the NV centre, might therefore serve as a novel versatile testbed for simulating different regimes of SR over a wide parameter range not easily accessible in other systems.

\color{Black}

\section*{Acknowledgments} We thank Marcus Doherty and Neil Manson for helpful discussions. This work was funded by the Australian Research Council Centre of Excellence for Engineered Quantum Systems (CE110001013). Comments or requests for materials should be addressed to: {\tt thomas.volz@mq.edu.au}

\section*{Methods}

\paragraph{Nanodiamond sample} The NDs used in this experiment are synthetic type Ib powders. The ND powder as received (MSY ${\leq}0.1$ $\mu \rm{m}$; Microdiamant) was used as the control sample to determine the average baseline value for the LT of single NV centres. Superradiance was investigated by using a second diamond powder which had been further treated to increase the concentration of NV centres. The NDs were purified by nitration in concentrated sulphuric and nitric acid (H2SO4-HNO3), rinsed in deionized water, irradiated by a 3-MeV proton beam at a dose of ($1{\times}10^6$ ions per $\rm{cm}^2$ and annealed in vacuum at $700 \,^\circ \rm{C}$ for 2 hours to induce the formation of NV centres (Academia Sinica, Taipei Taiwan \cite{fu_characterization_2007}). Both the as received and the irradiated NDs were characterized by means of a lab-built confocal scanning fluorescence microscope combined with a commercial atomic force microscope (AFM), described elsewhere \cite{bradac_prediction_2009}. For characterization, the diamond nanocrystals were dispersed on a $170$-$\mu \rm{m}$ thick BK7 glass coverslips (BB022022A1; Menzel-Glaser) which had been previously sonicated and rinsed in acetone (C$_3$H$_6$O, purity ${\ge} 99.5\% $; Sigma-Aldrich) for 10 min. The spectral interrogation of the NDs to identify emission from NV centres was performed via a commercial spectrometer (Acton 2500i, Camera Pixis100 model 7515-0001; Princeton Instruments). The size of each individual ND was measured using a commercial atomic force microscope (Ntegra; MT-NDT); the value for the average ND size is ($110 \pm 30$) $\rm{nm}$. While for the as received sample the concentration of NV centres is extremely low (at most a few NVs per nanocrystals), for the irradiated one we estimate a concentration of ${\sim} 3{\times}10^6$ NV centres per $\mu \rm{m}^3$. This was determined by correlating – for nanocrystals of different sizes – the average fluorescence intensity measured for each ND with its volume, and cross-checking this ratio with the one given by the sample provider \cite{fu_characterization_2007}.

\paragraph{Measurements} Lifetime measurements were performed under off-resonant laser excitation ($\lambda = 532\,\rm{nm}$), for which we employed a pulsed laser source (LDH-P-FA-530; PicoQuant) with the repetition rate set at either $5$ or $20\,\rm{MHz}$. Emission from the NV centres was filtered either via a long-pass filter (FEL0650, FEL0700; Thorlabs) or via a spectrometer (SpectraPro Monochromator Acton SP2500, dispersion $6.5\,\rm{nm/mm}$ at $435.8\,\rm{nm}$; Princeton Instruments) used as a monochromator and centred around the NVs' ZPL. The emitted photons were detected using an ID Quantique id100-20-ULN single-photon avalanche photodiode (APD). In order to achieve higher quantum efficiency, photon-coincidence measurements were performed using a set of two Perkin Elmer SPCM-AQR-14 APDs instead. They were arranged in a Hanbury-Brown and Twiss interferometer configuration (Fig. \ref{fig:Figure3}a) in order to determine the second-order correlation function $g^{(2)}(\tau)$. For pulsed excitation, we measured the time-integrated autocorrelation function, $\overline{g^{(2)}(\tau)} \equiv \int_{-\tau}^\tau dt  \langle :I(0) I(t):\rangle/ \int_{-\tau}^\tau dt \langle I(0) \rangle  \langle I(t)\rangle$, where $\langle I(t) \rangle$ is the luminescence signal intensity. This is evaluated by normalizing the photon coincidences of the '0' peak against the other peaks, for time slices of increasing duration.  As $\tau \rightarrow 0$, $\overline{g^{(2)}(\tau)}$ approaches the standard autocorrelation function $g^{(2)}(0)$, which was measured to identify single NV centres in the control ND sample displaying the characteristic photon anti-bunching dip signifying sub-Poissonian count statistics. On the other hand, superradiant NDs revealed super-Poissonian statistics characterized by photon bunching with a corresponding $\overline{g^{(2)}(0)} \ge 1$ and up to $1.14 \pm 0.02$, (cf. main text, Fig. \ref{fig:Figure3}b--d).

\bibliographystyle{naturemag} \bibliography{20160810_SRrefs}

\newpage

\pagebreak

\begin{center}
\textbf{\Large Supplementary Information}
\end{center}

\section{Theoretical model and fitting}

\begin{figure}[t!]
\includegraphics[width=6cm]{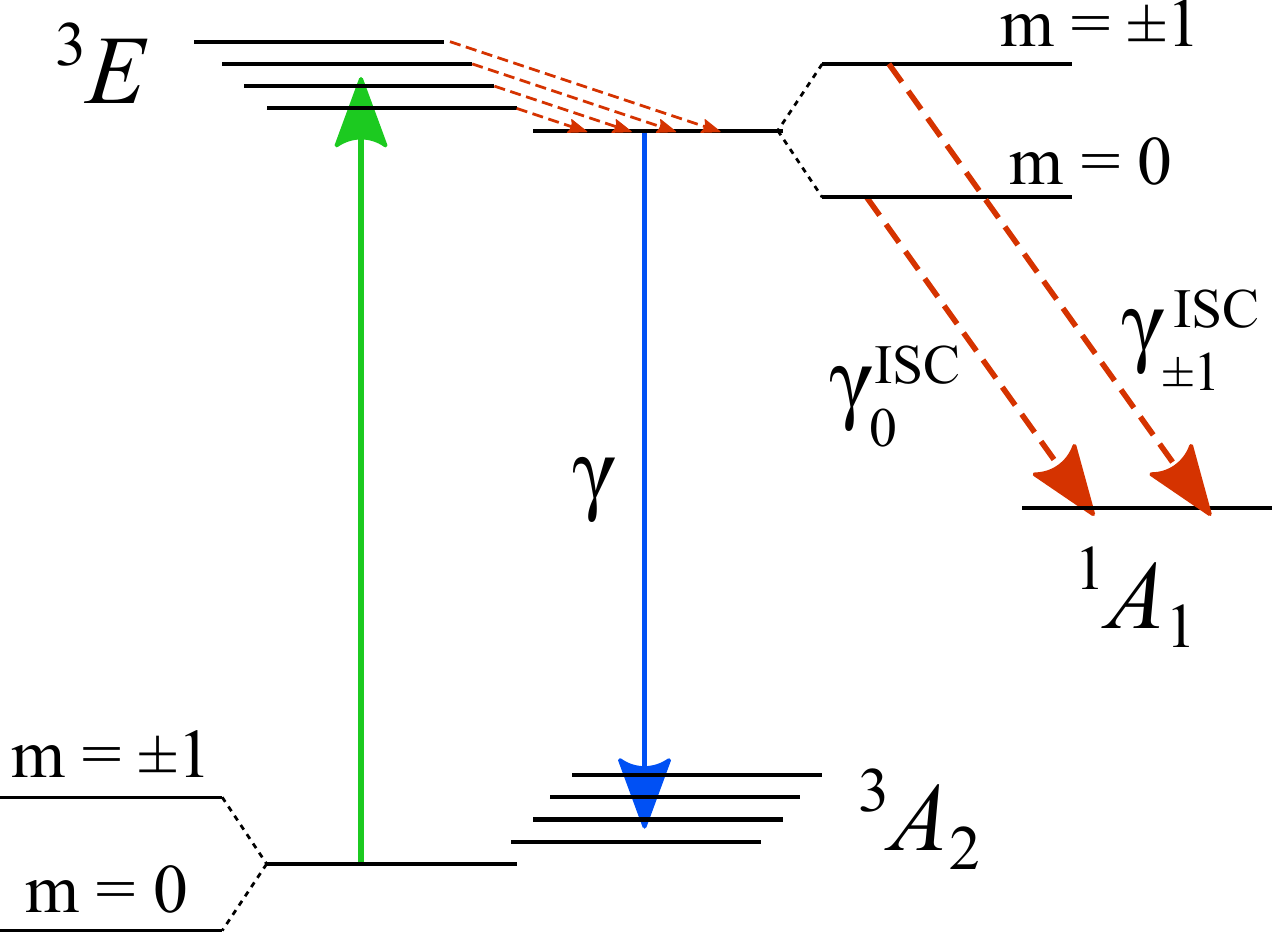}
\centering
\caption{NV centre level scheme used by the theoretical model, with solid lines depicting electromagnetic transitions and dashed lines show phononic transitions. Population in the ground state manifold ($^3 A_2$) is off-resonantly excited (green arrow) to a variety of excited states, which then relax down to the excited $^3 E$ manifold over a few tens of picoseconds. Initially this excited population consists entirely of domains that act collectively, but over time local dephasing can move portions of the collective populations into excited, but non-collective states. Population in the $^3 E$ manifold undergoes collective and non-collective electromagnetic decay to various phononic sidebands of the electronic ground state ($^3 A_2$). The excited manifold can also relax phononically to the $^1 A_1$ manifold via the inter-system crossing rate. As the experimental detection system filters out the ZPL, only photons that decay to the excited phononic sidebands of the electronic ground state are seen seen. Excitation and decays preserve the spin project quantum number $m\equiv\sigma$, so population in the $m=0$ and $m=\pm 1$ levels are treated independently and their fluorescence added to produce the final result. }
\label{fig_theory_level_scheme}
\end{figure}

In our model, each NV centre is treated as shown in Figure~\ref{fig_theory_level_scheme}. A green 532nm laser excites the system into one of several vibronic levels of the excited electronic state. Within a few tens of picoseconds the system relaxes down to the vibronic ground state of the excited electronic state. From here the centre can decay electromagnetically to the various phononic states of the electronic ground state. As we are working at room temperature, we expect about 3\% of the decay goes to the phononic ground state, i.e. the zero phonon line (ZPL), and 97\% of the decay goes into the various excited vibronic sidebands \cite{Santori:2010fk}.

The electronic excited state can also decay via dark, i.e. not mediated by the electromagnetic field, decay to the $^1A_1$ manifold, where it eventually relaxes back to the electronic ground state over a time $\sim 100$\,ns. These so called intersystem crossing (ISC) rates, $\gamma_{\sigma}^{\rm ISC}$, depend on the spin state $\sigma$ and while they are not directly observed, they can be determined from measured decay rates according to the formula \cite{doherty_nitrogen-vacancy_2013}:
\[
\frac{T_{\pm 1}}{T_{0}}=\frac{1+f_0}{1+f_1},
\]
where $T_{\sigma}$ is the total lifetime for excited state with spin projection $m\equiv\sigma$, $\ket{e_{\sigma}}\equiv \ket{^{3}E,\sigma}$, and $f_{\sigma}=\gamma_{\sigma}^{\rm ISC}/\gamma$ where $\gamma$ is the spin independent radiative decay rate \cite{PhysRevLett.100.077401}. The optical decay rate for bulk diamond is $\gamma=2\pi \times 12.2$\,MHz \cite{PhysRevB.74.104303} and from experiments in bulk diamond at room temperature \cite{PhysRevLett.100.077401, PhysRevLett.114.145502} the ISC rates are $\gamma^{\mathrm{ISC}}_{\pm 1} = 2\pi \times 9.4$\,MHz and $\gamma^{\mathrm{ISC}}_{0} = 2\pi \times 1.8$\,MHz. The ISC rates have not been measured in nanodiamond but in our model we assume the same values as in bulk.

We treat the ensemble of NV centres in each nanodiamond as one that can be divided into domains of various sizes. Within each domain $k$ there are $N_k$ centres that act collectively for some time after their initial excitation. We assume the electromagnetic decay from the $^3E$ manifold to the $^3A_2$ is collective, but the ISC decay is not. Each centre acts as a quantum mechanical spin and we label the relevant ground and excited states that participates in the spin dynamics as
\[
\ket{e_{\sigma}}\equiv \ket{^{3}E,\sigma},\quad \ket{g_{\sigma}}\equiv \ket{^{3}A,\sigma},
\]
where $\sigma$ is the electron spin projection.
To describe the collective behaviour within a domain we introduce the collective spin vector operators
\begin{equation}
\hat{\vec{S}}^{(\sigma)} = \sum ^N _{j=1} \hat{\vec{s}}_j ^{(\sigma)},
\end{equation}
where $\hat{\vec{s}}^{(\sigma)}_j$ is the single NV centre spin operator with spin polarization $\sigma$ at site $i$ in a domain with $N$ collective NV centres. Specifically, $\hat{\vec{s}}^{(\sigma)}_j=\hat{s}^{(\sigma)x}_j\hat{x}+\hat{s}^{(\sigma)y}_j\hat{y}+\hat{s}^{(\sigma)z}_j\hat{z},$ where $\hat{s}^{(\sigma)x}_j=(\hat{s}_j^{(\sigma)+}+\hat{s}_j^{(\sigma)-})/2, \hat{s}^{(\sigma)y}_j=(s_j^{(\sigma)+}-s_j^{(\sigma)-})/2i$ with the raising and lowering operators given by
\[
\hat{s}_j^{(\sigma)+}=\ket{e_{\sigma}}\bra{g_{\sigma}}=(\hat{s}_j^{(\sigma)-})^{\dagger},
\]
and the $\hat{z}$ component
$\hat{s}^{(\sigma)z}_i=(\ket{e_{\sigma}}\bra{e_{\sigma}}-\ket{g_{\sigma}}\bra{g_{\sigma}})/2$.
We will make use of the collective spin components 
\begin{equation}
\hat{S}^{(\sigma)\pm} = \sum ^N _{j=1} \hat{s}_j^{(\sigma)\pm},\quad \hat{S}^{(\sigma)z}=\sum ^N _{j=1} \hat{s}_j^{(\sigma)z}.
\end{equation}
in the description of the dynamics.

Because the transitions considered here are to a very good approximation spin conserving, the dynamics occurs in independent subspaces labelled by the spin projection $\sigma$. Hence we solve for the dynamics of the joint state $\hat{\rho}^{(\sigma)}$ within each subspace and then average based on the initial polarization of the spins.
We can write the Louivillian due purely to collective decay within each spin subspace as (henceforth we set $\hbar\equiv 1$)
\begin{equation}
\frac{d \hat{\rho}^{(\sigma)}}{dt} = \mathcal{L}[\hat{\rho}^{(\sigma)}],
\end{equation}
where
\[
\mathcal{L}[\hat{\rho}^{(\sigma)}]=\gamma \left[ 2\hat{S}^{(\sigma)-}\hat{\rho}^{(\sigma)} \hat{S}^{(\sigma)+}  - \{\hat{S}^{(\sigma)+}  \hat{S}^{(\sigma)-}, \hat{\rho}^{(\sigma)}\}\right].
\]
For nanodiamonds the optical decay rate $\gamma$ can be considerably slower than in bulk.  This rate reduction is due the reduced size affecting the density of states as well as a number of other factors such as surface effects. In our samples the optical decay rate ranged from $\gamma\sim 2\pi\times 3.2$\,MHz -- $2\pi\times  12.7$\,MHz. In the real environment of NV centres there are other processes not captured by collective decay which must be included in the model and which we describe below.

We denote the maximal spin of the collective ensemble by $J$, with $\hat{S}^{(\sigma)2} = J(J+1)$. As each individual NV centre is a spin-1/2 system, we have $J=N/2$ where $N$ is the total number of collective centres in our domain. We label the eigenstates of the collective spin operator $\hat{S}^{(\sigma)z}$
by the eigenvalues $M\in\{ -J,-J+1 \ldots J-1,J\}$. The maximal spin (or collective state) $|J,M,\sigma\rangle$ can be written as
\begin{equation}
|J,M,\sigma\rangle = \sqrt{\frac{(J+M)!(J-M)!}{(2J)!}} \sum_{perm} |\underbrace{e_{\sigma}\ldots e_{\sigma}}_{J+M} \,\, \underbrace{g_{\sigma}\ldots g_{\sigma}}_{J-M} \rangle,
\label{symstate}
\end{equation}
where the sum is over all permutations of the $N$ spins. 

In the following we will treat the $\sigma=\pm 1$ states as indistinguishable as, at high temperature, the spin orbit coupling in the excited state does not couple them so they are degenerate and couple equally to the environment  \cite{doherty_nitrogen-vacancy_2013}. This means that we can have Dicke states $|J,M,\sigma\pm 1\rangle $ as in Eq.~(\ref{symstate}) where the permutation does not distinguish between spin projection $+1$ and $-1$. Henceforth, to simplify notation we use the variable $\sigma$ to refer to the absolute value of spin projection: $\sigma=|m|$.

The initial state of our system, after excitation of the laser and after relaxation of excited vibronic states to the $^{3}E$ manifold optically excited state, is assumed to be a mixture over spin ensembles $\hat{\rho}^{(\sigma)}(0)$, each one being a mixture over Dicke states, i.e.
\begin{equation}
\hat{\rho}^{(\sigma)}(0)=\sum_{N_{\sigma}}p_{N_{\sigma}}\hat{\rho}^{(\sigma)}_{N_{\sigma}}(0),
\end{equation}
where $p_{N_{\sigma}}$ is the probability a domain of size $N_{\sigma}$ exists with spin $\sigma$. Note $\sum_{N_{\sigma}}p_{N_{\sigma}}=p_{\sigma}$, i.e. the total fraction of contributing spins in state $\sigma$, and the overall normalization is $\sum_{\sigma=0,1}p_{\sigma}=1$. The collective state is 
\begin{equation}
\hat{\rho}^{(\sigma)}_{N_{\sigma}}(0)=\sum_{M} P^{(\sigma)}_{J=\tfrac{N_{\sigma}}{2},M}(0) \ket{\tfrac{N_{\sigma}}{2},M,\sigma}\bra{\tfrac{N_{\sigma}}{2},M,\sigma},
\label{initstate}
\end{equation}
 i.e. a state diagonal in the maximum angular momentum subspace. The model was found to fit the fluorescence data by assuming a maximally mixed state in this subspace with occupation probabilities: $P^{(\sigma)}_{\tfrac{N_{\sigma}}{2},M}(0)=\frac{1}{N_{\sigma}+1}$. The justification for this form of the initial state is as follows. Before the excitation laser is turned on, the state of the spins within a domain with spin projection $\sigma$ is $\ket{\Psi}_1=\ket{J,-J,\sigma}$. The interaction of the spins with the laser is symmetric and prepares the coherent superposition $\ket{\Psi}_2=\sum_{M,n_{\nu}}c^{(\sigma)}_{M,n_{\nu}} |J,M,\sigma,n_{\nu}\rangle $ where $|J,M,\sigma,n_{\nu}\rangle$ is a symmetric state like in Eq.~(\ref{symstate}) but where the excited state is $\ket{e_{\sigma,n_{\nu}}}=\ket{^{3}E,n_{\nu},\sigma}$ where $n_{\nu}$ is the vibrational quantum number associated with a vibrational mode $\nu$ that the laser most strongly couples to. The coefficients $c^{(\sigma)}_{M,n_{\nu}}$ depend on the Rabi frequency, Franck-Condon factor, and detuning of the laser. In Ref. \cite{huxter_vibrational_2013} it was found that the vibronic mode frequencies of the spin triplet states of NV$^{-}$ defect occur in the range $5-46$ THz and hence are well separated in energy with respect to the laser coupling. These modes are local to each spin in the sense that they are associated with deformations in the lattice within a few unit cells of the defect which leads to a modification of the elastic moduli. The excited local vibronic states are short lived with lifetimes on the order of a few picoseconds \cite{huxter_vibrational_2013}. The nature of the relaxation is decay via coupling to local environmental modes which are in turn strongly coupled to long range phonon modes in the lattice. The combined effect is to erase the information about which spin experienced relaxation and hence the decay process preserves the permutation symmetry of the collective states so that $|J,M,\sigma,n_{\nu}\rangle \rightarrow |J,M,\sigma\rangle$ for any initially excited vibrational state. During this process, there are global dephasing processes due to non energy exchanging (dispersive) phonon couplings occurring at room temperature a rate of $\sim 1$ THz \cite{fu_observation_2009} which will damp coherences between collective states resulting in diagonal states of the form $\rho^{(\sigma)}(0)$ in Eq.~(\ref{initstate}). We assume the exactly evenly mixed state as it fit the experimentally observed fluorescence data and the $g^{(2)}(0)$ measurements presented in the main text. Small deviations from this initial statistical distribution do not significantly change the results.
 
In the dynamical evolution, to simplify the model, we assume that all population resides either within a collective subspace spanned by the states $\{\ket{J,M,\sigma}\}$ or as independent spins. We have a series of collective subspaces, with dimensions $2, 3, \ldots, 2J+1$ and if a subspace contains $N$ spins, its quantum number $J$ is given by $J=N/2$. Since we are interested in calculating fluorescense rates we need only track populations in the collective states $P^{(\sigma)}_{J,M}(t) = \langle J,M,\sigma | \hat{\rho}^{(\sigma)}(t) | J,M,\sigma \rangle$ or single spin excited states, and not coherences between them. Thus we need only solve rate equations and not the full master equation for the density matrix.
The rate equation including collective decay only is
\begin{equation}
\begin{split}
\frac{d P^{(\sigma)}_{J,M}(t)}{dt} = 2\gamma \big[ & (J(J+1) - M(M+1)) P^{(\sigma)}_{J,M+1}(t)  \\
                                     & - (J(J+1) - M(M-1)) P^{(\sigma)}_{J,M}(t) \big].
\end{split}
\end{equation}
This equation assumes that the collective subspace remains intact throughout the entire decay process. More realistically, there will be local decoherence processes that remove spins from the subspace. We account for two such mechanisms. The first is the aforementioned intersystem crossing decay at rate $\gamma_{\mathrm{ISC}}^{\sigma}$, which acts as local leakage to project any given spin from the excited state to the singlet state $^1A_1$ which is dark and does not contribute to the fluorescence signal. The second is a local dephasing+projection map which preserves excitation number but can couple a collective state $\ket{J,M,\sigma}$ to the state $\ket{e_{\sigma}}_j\otimes \ket{J-1/2,M-1/2,\sigma}$ which has one fewer spin in the collective space and one spin in the excited state. We describe this as a composite map occuring at some rate $\gamma_d$ involving local spin dephasing followed by local projection onto the excited state.
 A local unitary phase flip on the collective state is $2\hat{s}_j^{(\sigma)z}|J,M\rangle$ and the overlap with the collective state is
\begin{equation}
\langle J,M,\sigma |2\hat{s}_j^{(\sigma)z} | J,M,\sigma  \rangle = -M/J.
\end{equation}
This means any constituent excited state spin in the state $| J,M,\sigma  \rangle$ becomes decoupled from the collective subspace at a rate $\gamma^{\rm d}_{\sigma} |M/J|^2$ and joins the non-collective population where it will undergo normal exponential decay. The subspace itself has lost an atom, and is projected into another collective subspace of size $2J-1$ at the complementary rate $\gamma^{\rm d}_{\sigma} (1-|M/J|^2)$.

The rate equation for the subspace with spin angular momentum $J$ including local decoherence is 
\begin{equation}
\begin{array}{lll}
\frac{d}{dt} P^{(\sigma)}_{J,M}(t) &=& \gamma \Big[  (J(J+1) - M(M+1)) P^{(\sigma)}_{J,M+1}(t) \\
                        && \,\,\,\, - (J(J+1) - M(M-1)) P^{(\sigma)}_{J,M}(t) \big] \\
                        & &- 2 J \gamma^{\rm d}_{\sigma} (1 - \left| M/J \right|^2) P^{(\sigma)}_{J,M}(t) \\
                        & & + 2(J+\tfrac{1}{2}) \gamma^{\rm d}_{\sigma} \left( 1 - \left| \frac{M+\tfrac{1}{2}}{J+\tfrac{1}{2}} \right|^2 \right) P^{(\sigma)}_{J+\tfrac{1}{2},M+\tfrac{1}{2}}(t)\\
&& +\gamma^{\rm ISC}_{\sigma} \Big((J+M+1) P^{(\sigma)}_{J+\tfrac{1}{2},M+\tfrac{1}{2}}(t)  \\
&&- (J+M)  P^{(\sigma)}_{J,M}(t)  \Big).  
\label{fulldynamics}             
\end{array}
\end{equation}
In this form it is clear that our local dephasing+projection map causes a given collective subspace to gain population from the subspace one spin larger, and lose population to the subspace with one spin smaller. There is also global dephasing at a rate of $\sim 1$THz \cite{fu_observation_2009} due to coupling to long wavelength phonon modes, but because this only affects coherences between collective states this does not enter into the rate equations.
  
We also need keep track of the number of atoms decaying independently. The relevant quantity is the number of independent spins in the excited state, which we denote $N_{\mathrm{nc}}$. This number is fed by the dephasing+projection process acting on the collective subspace and depleted by the single spin decay which includes the non-radiative decay: 
\begin{equation}
\begin{array}{lll}
\frac{d}{dt} N^{(\sigma)}_{\mathrm{nc}}&=&-(\gamma+\gamma^{\rm ISC}_{\sigma}) N^{(\sigma)}_{\mathrm{nc}}\\
&&+\gamma^{\rm d}_{\sigma} \sum_{J=1/2}^{N/2} \sum_{M=-J}^J \left(1-\left|\frac{M}{J}\right|^2 \right) 2 J P^{(\sigma)}_{J,M}.
\end{array}
\end{equation}

For a given domain size $N_{\sigma}=2J$ and spin $\sigma$, the fluorescence rate is
\begin{equation}
\begin{array}{lll}
F_{N_{\sigma}}(t) &=&\gamma \left(N^{(\sigma)}_{\mathrm{nc}}(t) +{\mathrm{Tr}} \left[ \hat{S}^{(\sigma)+} \hat{S}^{(\sigma)-} \hat{\rho}^{(\sigma)}(t) \right] \right)\\
&=&\gamma \bigg(N^{(\sigma)}_{\mathrm{nc}}(t)\\
&&+\sum_{J=1/2}^{N/2}\sum_{M=-J}^J(J(J+1)-M(M+1))\\
&&P^{(\sigma)}_{J,M}(t) \bigg).
\end{array}
\label{FR}
\end{equation}
The total fluorescence is then obtained by a weighted sum over spin ensembles and domain sizes therein
\begin{equation}
F(t)=\sum_{N_\sigma}p_{N_{\sigma}}F_{N_{\sigma}}(t).
\end{equation}

In order to solve this set of rate equations we write the populations in each spin subspace as a vector $\vec{v}^{(\sigma)}(t)$ with $(N^2 +3N)/2$ entries, where $N$ is the number of atoms in described by the largest collective subspace. We then create a matrix $ A^{(\sigma)}$ describing the evolution of $\vec{v}^{(\sigma)}(t)$ so that
\begin{equation}
\frac{d}{dt}\vec{v}^{(\sigma)}(t) = A^{(\sigma)} \vec{v}^{(\sigma)}(t).
\label{eqMasterEquationVectorizedRho}
\end{equation}
As $A^{(\sigma)}$ is time-independent, Eq.~(\ref{eqMasterEquationVectorizedRho}) can be solved by simple matrix exponentiation, with a computational resourse cost of $N^6$ for each time point. In practice, however, scaling is considerably worse for $N>50$ because we begin to exceed cache space on our CPU. This scaling limits us to $\lesssim 70$ centres for a single simulation, or $\lesssim 50$ when fitting a multidimensional parameter space. These limits are more than adequate for almost all the diamonds in our ensemble, with most showing collective effects of $N \lesssim 10$.

Once we have this solution, the time-dependent fluorescence rate $F(t)$ is easily obtained using Eq.~(\ref{FR})  as
where $N_{\mathrm{nc}}(t)$ is the number of excited, noncollective atoms which are undergoing normal exponential decay at that time. This incoherent excited population is tracked through numerical integration of the number of atoms being projected out of the collective subspaces over time, along with losses due to the optical and ISC decay rates.

This final theory lifetime curve is then convolved with the experiment's detector response function, which has been determined to be 110ps based on a measurement with a 5ps laser pulse. 

We model the $\sigma=0$ and $\sigma=1$ populations separately and add their fluorescence
 contributions independently. We assume the ISC rates are the same as in bulk diamond,
 and choose the optical decay rate $\gamma$ by fitting to the long term tail of
 the lifetime fluorescence curve, when the collective effects have ended and only
 the non-collective electromagnetic and ISC decay channels remain.
 For the intial state we assume equal population in each of the $M$ eigenstates
 of the largest collective subspace.

In order to fit this model to the experimentally measured lifetime curves we need to choose the number of collective centres per domain. We assume that the domain sizes (i.e. the number of atoms acting collectively in each domain) have a Gaussian distribution, with a number $N$ in largest domain. The other possible free parameters are the degree of $m=0$ spin polarization given by $p_0 $ as well as the dephasing rates $\gamma^d_{0}$ and $\gamma^d_{\pm 1}$. When performing the fits, however, these parameters remained relatively constant across all the diamonds. Typically the degree of spin polarization was $\sim$50\% -- 60\%, in agreement with \cite{felton_hyperfine_2009,acosta_diamonds_2009}, and the dephasing rates ranged from $\gamma^d_{0} \sim 2\pi\times 20$\,MHz -- $2\pi\times 40$\,MHz and $\gamma^d_{\pm 1} \sim 2\pi\times 300$\,MHz -- $2\pi\times 450$\,MHz.

We fit parameters to dozens of NVs centres. For the four representative diamonds we have chosen to profile in the paper, (see Figure 2 in main manuscript), the following fit parameters were extracted:

\hspace{2mm}

\setlength{\tabcolsep}{1.5mm}

\begin{tabular}{c | c c c c c}
  \hline			
  Diamond & $N$ & $\gamma^d_{0} /2\pi$ & $\gamma^d_{\pm 1} /2\pi$ & $\gamma/2\pi$ & $p_0$ \\ \hline
  NV 1 & 2 & 27 MHz & 270 MHz & 2.5 MHz & 0.56 \\
  NV 2 & 7 & 20 MHz & 260 MHz & 4.8 MHz & 0.51 \\
  NV 3 & 10 & 39 MHz & 420 MHz & 3.3 MHz & 0.50 \\
  NV 4 & 50 & 20 MHz & 450 MHz & 7.9 MHz & 0.50 \\
  \hline  
\end{tabular}

\hspace{4mm}

The values of maximum domain size $N$ and polarization $p_0$ found to best fit the data were determined as follows. We defined two sets $S_0$ and $S_1$ consisting of collective domain sizes for spins with projection $\sigma=0$ or $\sigma=1$. These sets were chosen with a variable maximum domain size and other sizes symmetrically distributed about that maximum value according to a probability distribution fixed for all the diamonds. This means the only two adjustable parameter were the maximum domain size in each set. The total number of spins in each set are $N_{\sigma \rm tot}=\sum_{n\in S_{\sigma}}n$ and the maximum domain size is the largest element of both the sets $S_0$ and $S_1$. Because the samples were always somewhat polarized along $\sigma=0$, this maximum is $N=\max S_0$. Given these sets of possible domain sizes, the probability for a given domain size $N_{\sigma}$ is
$p_{N_{\sigma}}=\frac{\sum_{n\in S_{\sigma}}n\delta_{n,N_{\sigma}}}{N_{0 \rm tot}+N_{1 \rm tot}}$, and the polarization is calculated to be
\begin{equation}
p_{0}=\frac{\sum_{n\in S_0} n}{N_{0 \rm tot}+N_{1 \rm tot}}.
\end{equation}

Our model for local dephasing is phenomenological. To better understand the physical mechanism, consider the effect of optical dipole-dipole interactions. For a pair of NV centres, at positions $\vec{r}_1$ and $\vec{r}_2$, the individual spin conserving dipole-dipole interaction is (see e.g. \cite{Albrecht:2013uq})
\begin{equation}
\hat{H}_{\rm dd}=\sum_{\sigma_1,\sigma_2}V_{\rm dd}(\ket{e_{\sigma_1},g_{\sigma_2}}\bra{g_{\sigma_1},e_{\sigma_2}}+\ket{g_{\sigma_1},e_{\sigma_2}}\bra{e_{\sigma_1},g_{\sigma_2}})
\label{Vdd}
\end{equation}
where 
\begin{equation}
V_{\rm dd}=\frac{3\gamma b}{4(nk_0\Delta r)^3} (\hat{d}_1\cdot \hat{d}_2-3(\hat{d}_1\cdot \hat{n})(\hat{d}_2\cdot \hat{n})).
\end{equation}
Here $\hat{d}_j$ is the unit vector direction of the dipole $j$, $\Delta r=|\vec{r}_1-\vec{r}_2|$ is the separation, $\hat{n}=(\vec{r}_1-\vec{r}_2)/\Delta r$ is the unit vector separation between dipoles, $b\approx 0.03$ is the branching ratio to the ZPL \cite{Santori:2010fk}, $n=2.4$ is the index of refraction of the diamond crystal, $\gamma$ is the optical decay rate which we can take as $\gamma/2\pi=5$ MHz for the present estimate, and $k_0=2\pi/\lambda_0$ is the vacuum wavevector of the optical transition at wavelength $\lambda_0=639$nm. 
Substituting these values, the interaction strength is
\begin{equation}
V_{\rm dd}=2\pi \times 8.56 {\rm MHz}\times \left(\frac{10 {\rm nm}}{\Delta r}\right)^3 (\hat{d}_1\cdot \hat{d}_2-3(\hat{d}_1\cdot \hat{n})(\hat{d}_2\cdot \hat{n})).\\
\end{equation}
From the manufacturer provided numbers, the density of NV centres in the nanodiamonds is $\rho_{\rm NV} =10^{24}{\rm m}^{-3}$, which implies the mean separation between any pair is $\Delta r\approx 12$nm, meaning at this mean separation we expect a maximum interaction strength of $V_{\rm dd}\approx 2\pi \times  10$MHz.

From the form of the dipole-dipole interaction Eq.~(\ref{Vdd}), it clearly preserves the number of excitations, but is not permutation symmetric over a domain (and cannot be except for a collective space of size two, or a symmetrically arranged configuration of size three). 
Hence the interaction will act to couple collective states with different spin angular momentum $J$ while preserving the total excitation number $M$ quantum number. Unlike our model, $\hat{H}_{\rm dd}$ can couple to both higher and lower $J$ values but importantly it has a value which is no larger in magnitude to our inferred local dephasing rates $\gamma^0_d$, even after summing over dipole pairs at different separations and angles for our domain sizes, implying that at least we are not underestimating the effects of dipole-dipole interactions. Our inferred dephasing rates $\gamma^{\pm 1}_d$ are an order of magnitude larger. This is likely due to coupling to local electric fields at each spin, which has a larger coupling strength due to the significant excited state permanent dipole moment for the $\pm 1$ states \cite{doherty_nitrogen-vacancy_2013}.
 
To compare our model to non-collective behaviour, we attempted to fit the observed lifetimes with both a bi-exponential and a deformed exponential lifetime curve \cite{gatto_monticone_systematic_2013}. The fits using the bi-exponential model were consistently worse than our model with the discrepancy increasing as the collectivity increased. For the deformed exponential we used a coefficient appropriate to dipole-dipole coupling and a dimensionless coupling strength parameter \cite{gatto_monticone_systematic_2013} consistent with the defect density of our samples. This also resulted in poorer fits than our model with ratio of least squares errors up to 40 times worse.

\section{Second-order coherence}
The normalized second order coherence function for delay time $t$ and for a ensemble with domain size $N_{\sigma}$ is
\begin{equation}
g_{N_{\sigma}}^{(2)}(t) = \frac{\langle \hat{S}^{(\sigma)+}(0)  \hat{S}^{(\sigma)+}(t)  \hat{S}^{(\sigma)-}(t)  \hat{S}^{(\sigma)-} (0) \rangle} {\langle \hat{S}^{(\sigma)+}(0)  \hat{S}^{(\sigma)-}(0)\rangle\langle   \hat{S}^{(\sigma)+}(t)  \hat{S}^{(\sigma)-} (t) \rangle}.
\end{equation}
The total observed second order coherence is the sum
\begin{equation}
g^{(2)}(t)=\sum_{\sigma}\sum_{N_{\sigma}}p_{N_{\sigma}}g_{N_{\sigma}}^{(2)}(t). 
\end{equation}

If we assume all the population is initially in the largest collective subspace, then immediately after the excitation process is complete and before decay begins we have (here $J=N_{\sigma}/2)$
\begin{equation}
\begin{array}{lll}
{\mathrm{Tr}} \big[ \hat{S}^{+}  \hat{S}^{+}  \hat{S}^{-}  \hat{S}^{-}  \hat{\rho}_{N}(0) \big] 
&=& \sum_{M=-J} ^{J} P_{J,M}(0) (J(J+1)\\
&& - M(M-1)) \times (J(J+1) \\
                                     && - (M-1)(M-2)), \\
{\mathrm{Tr}} \big[ \hat{S}^{+} \hat{S}^{-} \hat{\rho}_{N}(0) \big] &=& \sum_{M=-J} ^{J} P_{J,M}(0) (J(J+1) \\
&&- M(M-1)).
\end{array}
\end{equation}
For zero delay, the result for $g^{(2)}(0)$ is clearly dependent on the choice of initial state. In the extreme case where the initial state consists of all spins up (corresponding to all atoms in the collective subspace being excited), we have $P^{(\sigma)}_{J,M} = \delta_{J,M}$, so that
\begin{equation}
g_{N_{\sigma}}^{(2)}(0) = 2 - \frac{2}{N_{\sigma}}. 
\end{equation}
Thus, for large domain sizes, the second order coherence function asymptotes to 2.

As described above, however, in our modelling we assumed an initial state where all $M$ eigenstates are equally populated. That is,
\begin{equation}
P^{(\sigma)}_{J,M}(0) = \frac{\delta_{J,N/2}}{2J+1} \,\, \forall \, M \in \{ -J, \ldots, J\}. 
\end{equation}
With this initial state we find
\begin{equation}
g_{N_{\sigma}}^{(2)}(0) = \frac{6 (N_{\sigma}-1)(N_{\sigma}+3)} {5 N_{\sigma} (N_{\sigma}+2) }, 
\end{equation}
which asyptotes to $g_{N_{\sigma}}^{(2)}(0) = 1.2$ as $N_{\sigma} \rightarrow \infty$.
For a set of varying domain sizes $N_{\sigma}$ with probability distribution $p_{N_{\sigma}}$, and initial states given by Eq.~(\ref{initstate}), the second order coherence is
\begin{equation}
g^{(2)}(0) = \sum_{\sigma}\sum_{N_{\sigma}}p_{N_{\sigma}} \frac{6 (N_{\sigma}-1)(N_{\sigma}+3)} {5 N_{\sigma} (N_{\sigma}+2) }. 
\label{g2zero}
\end{equation}
The solid red line plot in Fig. 3e of the main text plots $g^{(2)}(0)$ from Eq.~(\ref{g2zero}). Since we are mainly interested in the scaling with domain size, for simplicity we assume a single spin ensemble $p_{\sigma}=\delta_{\sigma,0}$ and $p_{N}$ Gaussian distributed with mean $\bar{N}$ given by the number of cooperative NVs (abscissa coordinate) and variance $\bar{N}/2$.  

\begin{figure}[htb!]
\includegraphics[width=8cm]{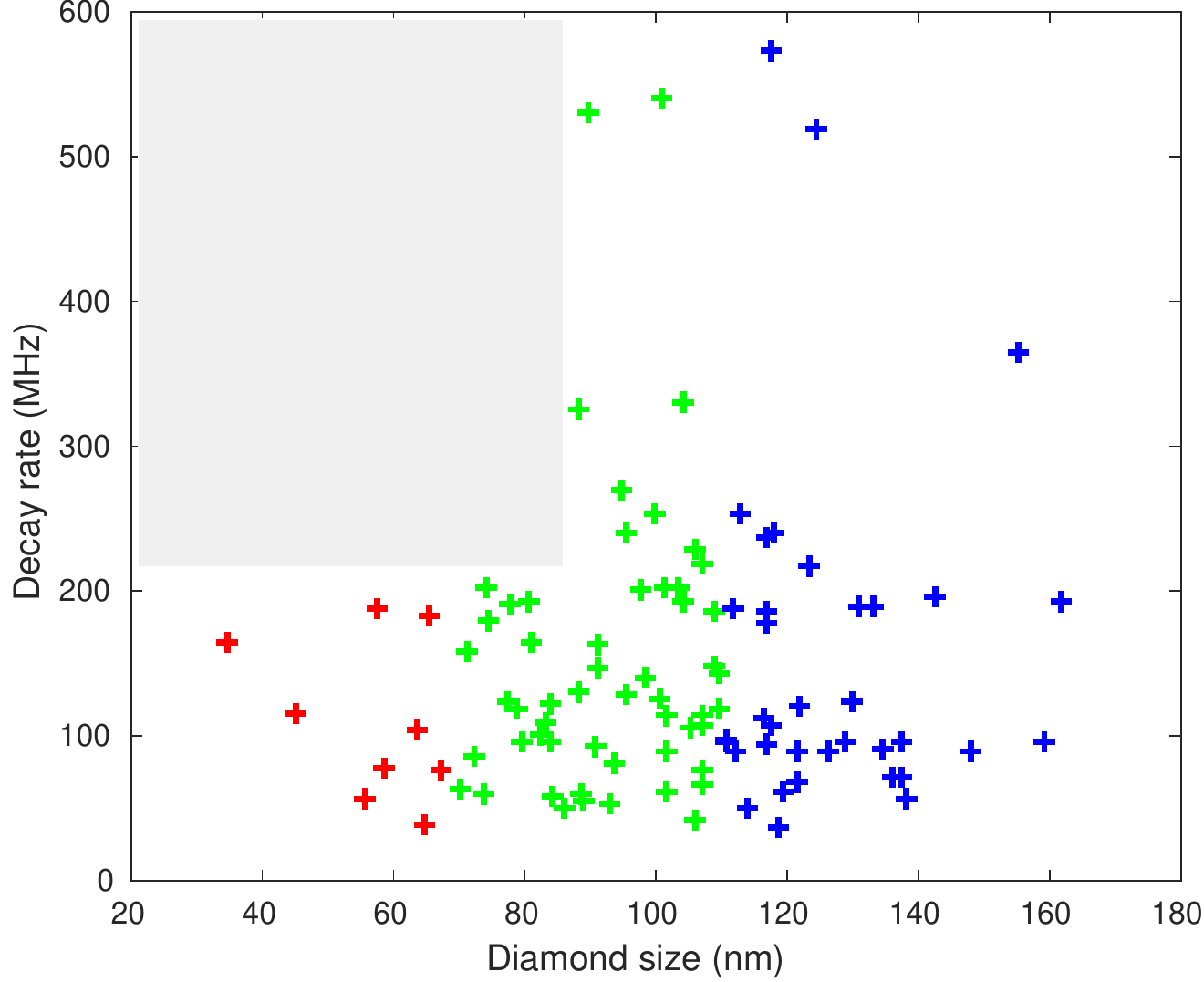}
\centering
\caption{The decay rate (reciprocal lifetime) for the 100 diamonds in the experimental ensemble as a function of diamond diameter. There is a forbidden region (shaded) showing that below a certain size, decays cannot be fast. This is consistent with our model of collective decays, and inconsistent with a model predicting that the fast decays arise from lattice defects, as such a model is independent of size. Color coding for nanodiamond diameter: red $<$ 70nm,  70nm $<$ green $<$ 110nm, blue $>$ 100nm}
\label{fig_decay_rate_vs_size}
\end{figure}

\begin{figure}[htb!]
\includegraphics[width=8cm]{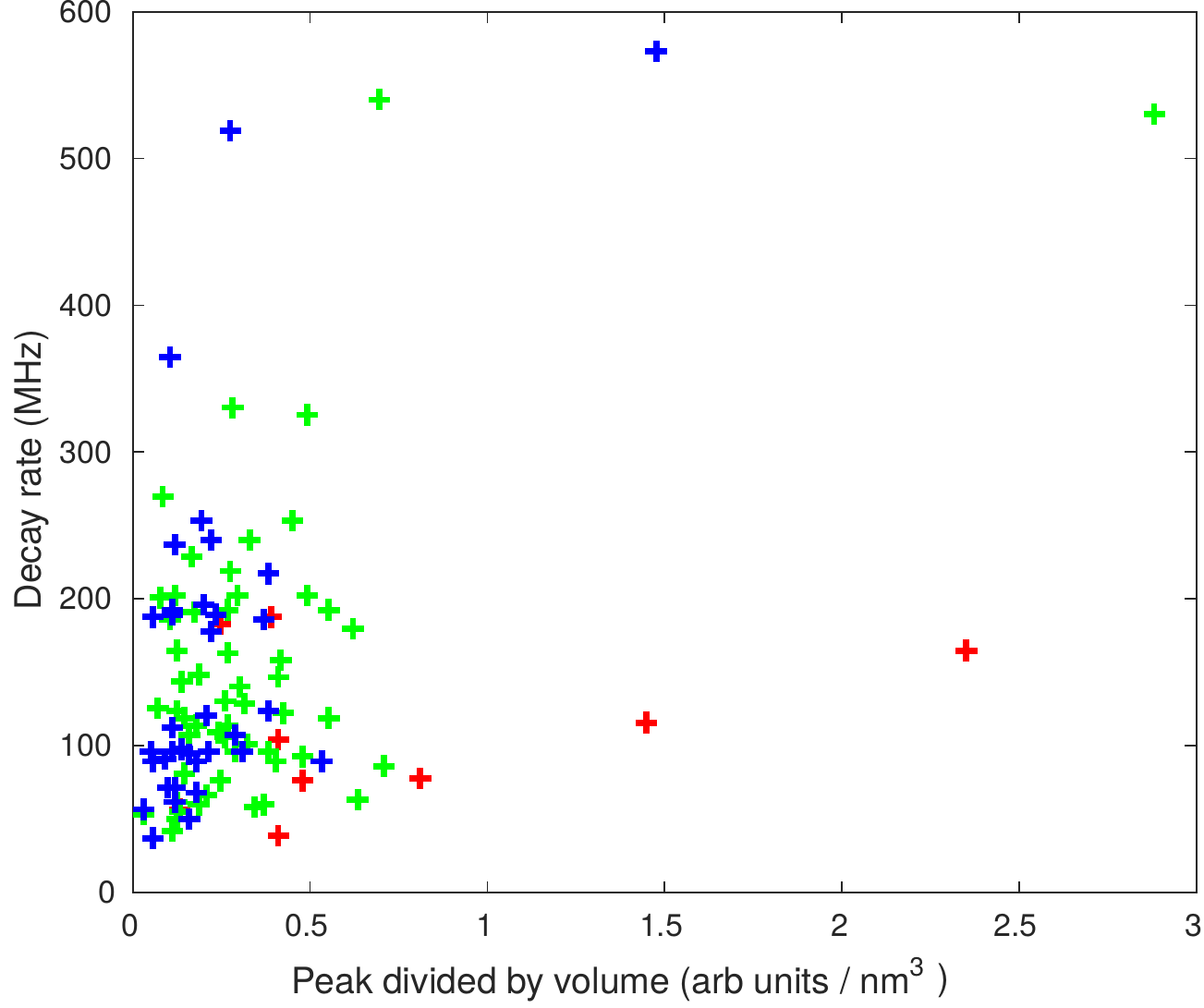}
\centering
\caption{The decay rate for the 100 diamonds in the experimental ensemble as a function of peak brightness during the lifetime measurement. Brightness has been normalized to diamond volume to account for the fact that larger diamonds have more emitters and are intrinsically brighter. The fact faster diamonds are not less bright disagrees with a lattice defect model (which predicts dark channel decay), and supports our collective decay model. Color coding for nanodiamond diameter: red $<$ 70nm,  70nm $<$ green $<$ 110nm, blue $>$ 100nm }
\label{fig_decay_rate_vs_brightness}
\end{figure}

\section{NV ensemble statistics}

As noted in the main manuscript, previous studies reported a decrease in the lifetimes of NVs for centres produced via low-energy He-ion radiation, with the decay time decreasing for increasing ion doses. This effect has been attributed to increased damage in the crystal lattice which provided nonradiative decay paths with faster dynamics \cite{mccloskey_helium_2014,orwa_raman_2000}, suggesting that the shortening of the lifetimes was due to nonradiative, `dark' pathways. In order to demonstrate that is not the relevant decay mechanism in our experiment, and that the lifetime shortening in our system is due to coherent collective effects, we took the entire ensemble of 100 nanodiamonds and examined the relationship between decay rates and size, brightness and NV centre density.

\begin{figure}[htb!]
\includegraphics[width=8cm]{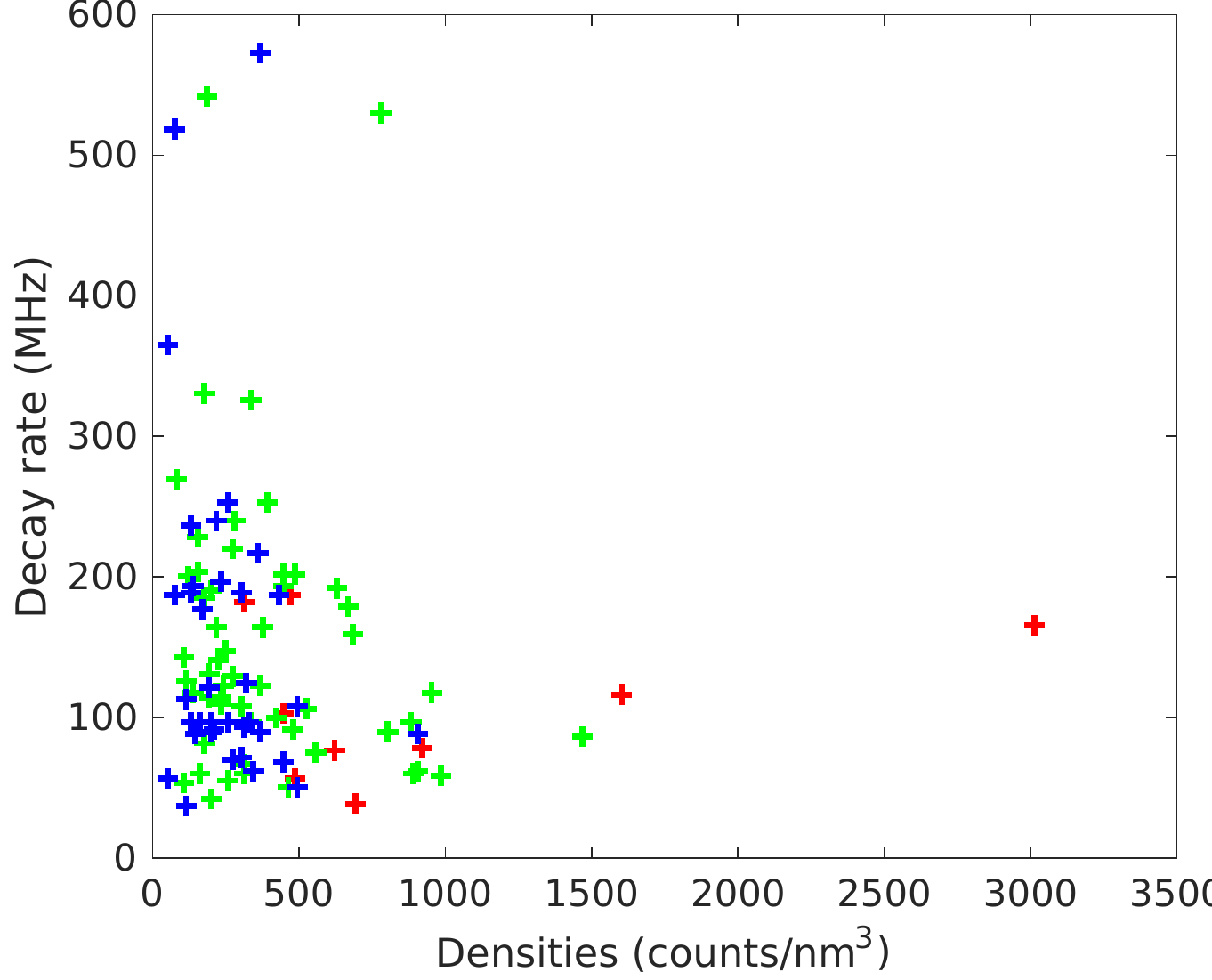}
\centering
\caption{The decay rate (reciprocal lifetime) for the 100 diamonds in the experimental ensemble as a function of the density of NV centres in each nanodiamond. Densities were determined by continuous fluorescence measurements. As the quantum efficiencies of the system are unknown, the densities are in arbitrary units, but their relative values are accurate. The lack of fast diamonds at high centre densities is consistent with our model of local dephasing arising from dipole-dipole interactions destroying collective effects. Color coding for nanodiamond diameter: red $<$ 70nm,  70nm $<$ green $<$ 110nm, blue $>$ 100nm }
\vspace*{2.4 in}
\label{fig_decay_rate_vs_densities}
\end{figure}

Figure~\ref{fig_decay_rate_vs_size} shows the relation between nanodiamond size and the decay rate, with each point on the plot corresponding to a single diamond in the ensemble. In order to not rely on a specific model for the analysis of the decay curves, we extracted the initial slope of the curve by fitting an exponential decay of the form $\exp [-t/\tau]$ to the first three nanoseconds of the lifetime curve, and then defining the rate as $1/\tau$. The lifetime curves do not follow standard exponential decay, especially for those with strong collective effects, but this method does produce a rate that reflects relative decay rates at short times.

The shaded region in the top left of Figure~\ref{fig_decay_rate_vs_size} contains no experimental data points indicating that small nanodiamonds ($<$70nm in diameter) cannot be fast. This is clearly inconsistent with hypothesis of crystal damage to lattice causing the fast decays. If it were true, then one would expect the decay speed up to be insensitive to diamond size --- the dark channels would be present for small diamonds as well as large diamonds. The lack of fast diamonds at small sizes is consistent with surface effects. As diamonds get smaller, the surface to volume ratio increases, and surface effects can break distinguishability, reducing the size of the collective domains. Thus Figure~\ref{fig_decay_rate_vs_size} is consistent with superradiant decay due to collective effect.

This is also confirmed by Figure~\ref{fig_decay_rate_vs_brightness}, which shows decay rate versus the peak fluorescence of the lifetime curve normalized to diamond volume (accounting for the fact that larger diamonds are intrisically brighter). The crystal damage hypothesis would suggest that faster diamonds should be less bright, since the acceleration of decay rate is due to dark channels. Again, no such effect is seen. If anything, a trend towards brighter diamonds being faster is seen, which is consistent with our model of collectively enhanced decay through a bright channel.

Finally we plot the decay rates versus NV centre density in Figure~\ref{fig_decay_rate_vs_densities}. The data suggests a trend that beyond a certain NV centre density, no rapid decay occurs. This is consistent with the idea that a high centre density leads to larger local dephasing due to dipole-dipole interactions.

\end{document}